\documentclass[utf8]{FrontiersinHarvard} 

\usepackage{url,hyperref,microtype,subcaption}
\usepackage[onehalfspacing]{setspace}
\usepackage{multirow}

\def\keyFont{\fontsize{8}{11}\helveticabold }
\def\firstAuthorLast{Guastavino {et~al.}} 
\def\Authors{Sabrina Guastavino\,$^{1,*}$, Francesco Marchetti\,$^{2}$, Federico Benvenuto\,$^{1}$, Cristina Campi\,$^{1}$ and Michele Piana\,$^{1,3}$}
\def\Address{$^{1}$MIDA, Dipartimento di Matematica, Università di Genova, Genova Italy \\
$^{2}$Dipartimento di Matematica "Tullio Levi Civita", Università di Padova, Padova, Italy \\
$^{3}$INAF - Osservatorio Astrofisico di Torino, Torino, Italy
}
\def\corrAuthor{Sabrina Guastavino}

\def\corrEmail{guastavino@dima.unige.it}

\begin{document}
\onecolumn
\firstpage{1}

\title[Operational flare forecasting]{Operational solar flare forecasting via video-based deep learning} 

\author[\firstAuthorLast ]{\Authors} 
\address{} 
\correspondance{} 

\extraAuth{}

\maketitle

\begin{abstract}

Operational flare forecasting aims at providing predictions that can be used to make decisions, typically at a daily scale, about the space weather impacts of flare occurrence. This study shows that video-based deep learning can be used for operational purposes when the training and validation sets used for the network optimization are generated while accounting for the periodicity of the solar cycle. Specifically, the paper describes an algorithm that can be applied to build up sets of active regions that are balanced according to the flare class rates associated to a specific cycle phase. These sets are used to train and validate a Long-term Recurrent Convolutional Network made of a combination of a convolutional neural network and a Long-Short Memory network. The reliability of this approach is assessed in the case of two prediction windows containing the solar storm of March 2015 and September 2017, respectively.


\tiny
 \keyFont{ \section{Keywords:} solar flares, deep learning, evaluation metrics, operational forecasting, machine learning} 
\end{abstract}

\section{Introduction}

Solar flare prediction is an important task in the context of space weather research, as it has to address open problems in both solar physics and operational forecasting \citep{schwenn2006space,2010AdSpR..45.1067M}. Although it is well-established that solar flares are a consequence of reconnection and reconfiguration of magnetic field lines high in the solar corona \citep{shibata1996new,sui2004evidence,su2013imaging}, yet there is still no agreement about the physical model that better explains the sudden magnetic energy release and the resulting acceleration mechanisms \citep{aschwanden2008keynote,shibata1996new,sui2004evidence,su2013imaging}. Further, solar flares are the main trigger of other space weather phenomena, and it is a challenging forecasting issue to predict the chain of the events that from solar flares lead to possible significant impacts on both in-orbit and on-Earth assets \citep{crown2012validation,murray2017flare}.

Flare forecasting rely on both statistical \citep{song_statistical_2009, Mason_2010, bloomfield2012toward, Barnes_2016} and deterministic \citep{strugarek2014predictive,petrakou2018deterministic} methods. In the last decade machine and deep learning algorithms have been obtaining an increasing interest, thanks to flexible algorithms that may take as input point-in-time feature sets extracted from magnetograms, time series of features, point-in-time images of active regions, and videos whose frames are made of magnetograms \citep{2021JSWSC..11...39G,2020ApJ...899..150N,2021EP&S...73...64N,guastavino2022implementation,bobra2015solar,liu2017predicting,nishizuka2018deep,florios2018forecasting,campi2019feature,li2020predicting,liu2019predicting,Sun_2022,pandey2022towards}. However, \cite{guastavino2022implementation} pointed out that the prediction performances of these supervised approaches are characterized by a notable degree of heterogeneity, which is probably related to significant differences in the way datasets are generated for training and validation. That study introduced an original procedure for the generation of well-balanced training and validation sets and discussed its performances by means of a video-based deep learning approach that combined a Convolutional neural Network (CNN)
with a Long Short-Term Memory (LSTM) network \citep{hochreiter1997long}. 

The present paper shows how that procedure can be used to build up an operational flare forecasting system that accounts for the periodicity of the solar activity. Specifically, applications are concerned with two temporal windows of the descent phase of Solar Cycle 24, comprising the "San Patrick's Storm" occurred in March 2015 \citep{astafyeva2015ionospheric,wu2016first,nayak2016peculiar} and the September 2017 storm \citep{campi2020machine,qian2019solar,Guastavino2019Desaturating}, respectively. Further, we assessed the prediction accuracy by using both standard skill scores like the True Skill Statistic (TSS) and the Heidke Skill Statistic (HSS), and the value-weighted skill scores introduced by \cite{guastavino2022bad} and better accounting for the intrinsic dynamical nature of forecasting problems \cite{guastavino2021prediction,hu2022probabilistic}. Results show that the use of the value-weighted version of TSS leads to predictions whose accuracy is in one case comparable to and in the other case much better than the one provided by the standard TSS.

The plan of the paper is as follows: Section 2 describes the data used for the analysis, the design of the neural network applied for the prediction, and the way operational flare forecasting is realized to account for solar cyclicity. Section 3 shows the results of the study. Our conclusions are offered in Section 4.

\section{Material and methods}

\subsection{The dataset}\label{sec:samples}

The archive of the Helioseismic and Magnetic Imager (HMI) \cite{scherrer2012helioseismic} on-board the Solar Dynamics Observatory (SDO) \cite{pesnell2011solar} contains two-dimensional magnetograms of continuous intensity, of the full three-component magnetic field vector and of the line-of-sight magnetic intensity. In our study, we considered the Near Realtime Space Weather {\em{HMI}} Archive Patch (SHARP) data products \cite{bobra2014helioseismic} associated to the line-of-sight components. Given an Active Region (AR), a sample associated to its history is a $24$-hour-long video made of $40$ SHARP images of an AR, with $36$ minutes cadence and where each image has been resized to a $128 \times 128$ pixels dimension. 

As in \cite{guastavino2022implementation} we defined seven different types of videos. Denoting the X1+, M1+, and C1+ the classes of flares with class X1 or above, M1 or above, and C1 or above, respectively, we have that
\begin{itemize}
\item X class samples are made of videos of ARs that originated X1+ flares in the next 24 hours after the sample time.
\item M class samples are made of videos of ARs that, in the next 24 hours after the sample time, generated flares with class M1+ but below class X.
\item C class samples are made of videos of ARs that, in the next 24 hours after the sample time, generated flares with class C1+ but below class M.
\item NO1 class samples are made of videos of ARs that never originated a C1+ flare.
\item NO2 class samples contain videos of ARs that originated neither a C1+ flare in the past nor in the next 24 hours after the sample time, but did originated it in the future.
\item NO3 class samples contain videos of ARs that did not originate a C1+ flare in the next 24 hours after the sample time, but did originate a C1+ flare in the 48 hours before the sample time (this definition accounts for the fact that relevant features like the flare index past and the flare past refer to the past 24 hour window \cite{campi2019feature}).
\item NO4 class samples contain videos of ARs that originated neither a C1+ flare in the next 24 hours after the sample nor a C1+ flare in the 48 hours before the sample time, but did originate a C1+ flare before the 48 hours before the sample time.
\end{itemize}

The first three types of samples describe the ability of ARs to generate flares of a given intensity in the next 24 hours (which is the prediction time interval), whereas the last four types of videos are associated to ARs which did not generate significant flares in the same prediction interval. However, samples labeled with 0 may not represent a quiescent situation: by instance, NO3 class samples are associated to ARs that generated intense flares during the observation period. Therefore, this type of NO samples are really hard to be distinguished from positive samples, which motivates the fact that they are often excluded from the analysis \citep{Sun_2022}. 

\subsection{Neural network architecture}
We used a deep neural network (DNN) which is appropriate for video classification. The DNN is called Long-term Recurrent Convolutional Network (LRCN) \cite{Donahue2017LRCN}: it is the combination of a convolutional neural network (CNN) and a recurrent neural network called Long-Short Term Memory (LSTM) network. The architecture is the same used in \cite{guastavino2022implementation} and it is summarized in Table \ref{tab:network}. Specifically, the CNN is characterized by the first four convolutional blocks with the number of nodes, the kernel size, the height and width strides and the activation function as input parameters. Each convolutional layer is $L_2$-regularized with the regularization level equal to $0.1$. The output of the last max-pooling layer is flatten and given in input to a fully connected layer of $64$ units, where dropout is applied with a fraction of $0.1$ input units dropped. Therefore, the output of the CNNs is a time series of $64$ features, which are
then passed to the LSTM which consists of $50$ units and where dropout is applied with a fraction of $0.5$ active units. 
Finally, the output of the LSTM layer is fed into the last fully connected layer, and the sigmoid activation function is applied to generate the probability distribution of the positive class, in order to perform binary classification.
The LRCN is trained over $100$ epochs using
the Adam Optimizer \citep{Kingma14} 
and mini-batch size equal to $128$.
    
\begin{table*}[ht]
		\centering
		\caption{Details of the LRCN architecture. The parameters of each layer, i.e. the number of filters, the kernel size, the height and width strides and the activation functions are shown.
}\label{tab:network}
{
\begin{tabular}{ | c | c c c c |}
\hline
Layer & Number of nodes & Kernel size & Stride & Activation function  \\
\hline  \\[-8pt]
Convolution & $32$ & $7\times 7$ & $2\times 2$ & Relu  \\
Batch Normalization & / & / & / & /  \\
Max pooling & / & $2\times 2$ & $2\times 2$ & / 
 \\
 \hline 
Convolution & $32$ & $5\times 5$ & $2\times 2$ & Relu  \\
Batch Normalization & / & / & / & /  \\
Max pooling & / & $2\times 2$ & $2\times 2$ & /   \\
\hline
Convolution & $32$ & $3\times 3$ & $2\times 2$ & Relu  \\
Batch Normalization & / & / & / & / \\
Max pooling & / & $2\times 2$ & $2\times 2$ & / \\
\hline
Convolution & $32$ & $3\times 3$ & $2\times 2$ & Relu \\

Batch Normalization & / & / & / & / \\
Max pooling & / & $2\times 2$ & $2\times 2$ & /   \\

Fully connected & $64$ & / & / & Relu  \\
\hline
LSTM & $50$ & / & / & /  \\
Fully connected & $1$ & / & / & Sigmoid \\
\hline
\end{tabular}
}
\end{table*}

\subsection{Loss functions and skill scores}
Loss functions and skill scores are intertwined concepts. On the one hand, in the training phase loss functions should be chosen according to the learning task \citep{rosasco2004loss}; on the other hand, in the validation/testing phase skill scores should account for the properties of the training set and the overall nature of the learning problem. In the classification setting, skill scores are usually derived from the elements of the so-called \textit{confusion matrix} (CM)
\begin{equation}\label{confusion-matrix}
\textrm{CM}= \left( \begin{array}{cc} 
\textrm{TN} & \textrm{FP} \\ 
\textrm{FN} & \textrm{TP}
\end{array}
\right),
\end{equation}
where the entries are the classical True Negative (TN), True Positive (TP), False Positive (FP) and False Negative (FN) elements. Score Oriented Loss (SOL) functions were proposed in \citep{marchetti2021score} and applied in \cite{guastavino2022implementation} for flare classification tasks. The concept at the basis of a SOL function is that it is defined starting from the definition of a skill score in such a way that the network optimization realized by means of that SOL function leads to the maximization of the corresponding skill score. 

SOL functions are constructed by considering a probabilistic version $\overline{\textrm{CM}}$ of the classical confusion matrix, which depends on a chosen cumulative density function (cdf) on $[0,1]$, (in this application the cdf is chosen as the one related to the uniform distribution). If $(f_w(\mathbf{x}_i),\mathbf{y}_i)$, $i=1,\dots,n$, are prediction-label couples of the classification task, where $f_w(\mathbf{x}_i)\in (0,1)$ is the probability outcome of the DNN $f_w$ on the sample $\mathbf{x}_i$ and  $\mathbf{y}_i\in\{0,1\}$ is the actual label associated to the sample $\mathbf{x}_i$, the $\overline{\textrm{CM}}$ entries are defined as
\begin{equation}
\begin{array}{c}
    \overline{\textrm{TP}}=\sum_{i=1}^n \mathbf{y}_i f_{w}(\mathbf{x}_i),\quad
    \overline{\textrm{TN}}=\sum_{i=1}^n (1-\mathbf{y}_i) (1-f_{w}(\mathbf{x}_i))\\
    \overline{\textrm{FP}}=\sum_{i=1}^n (1-\mathbf{y}_i) f_{w}(\mathbf{x}_i),\quad 
    \overline{\textrm{FN}}=\sum_{i=1}^n \mathbf{y}_i(1- f_{w}(\mathbf{x}_i)).
\end{array}
\end{equation}

In the space weather community, True Skill Statistics (TSS) is a relevant score, which is defined as 
\begin{equation}\label{eq:TSS}
        \textrm{TSS}(\textrm{CM})=\frac{\textrm{TP}}{\textrm{TP}+\textrm{FN}}+\frac{\textrm{TN}}{\textrm{TN}+\textrm{FP}}-1,
\end{equation}
and which is appropriate in classification problems characterized by class imbalance (TSS ranges in $[-1, 1]$ and it is optimal when it is equal to $1$). The SOL function associated to TSS is defined as
\begin{equation}\label{TTS-loss}
\ell_{\mathrm{TSS}}:= - \textrm{TSS}(\overline{\textrm{CM}}),
\end{equation}
which is a differentiable function with respect to the weights of the network and therefore it is eligible for use in the training phase. 

In the applications considered in this paper the prediction accuracy has been assessed by means of both TSS and the Heidke Skill Score (HSS)
\begin{equation}
    \textrm{HSS} =  \dfrac{2(\textrm{TP}\cdot\textrm{TN}-\textrm{FN}\cdot\textrm{FP})}{\textrm{P}\cdot(\textrm{FN}+\textrm{TN})+\textrm{N}\cdot (\textrm{TP+FP})},
\end{equation}
where $\textrm{P} = \textrm{TP}+\textrm{FN}$ and $\textrm{N} = \textrm{TN} + \textrm{FP}$. The HSS measures the improvement of forecast over random forecast, ranges in $(-\infty, 1]$,
and is optimal when it is equal to $1$.
Further, we considered the value-weighted skill scores introduced in \cite{guastavino2022bad}. These scores are based on a definition of confusion matrix that assigns different weights to FPs (denoted by wFPs) and FNs (denoted by wFNs) in such a way to account for the distribution of predictions along time with respect to the actual occurrences. By denoting the value-weighted confusion matrix as
\begin{equation}\label{confusion-matrix}
\textrm{wCM}= \left( \begin{array}{cc} 
\textrm{TN} & \textrm{wFP} \\ 
\textrm{wFN} & \textrm{TP}
\end{array}
\right),
\end{equation}
predictions are assessed by computing the value-weighted TSS (wTSS) and value-weighted HSS (wHSS), defined as follows
\begin{equation}
    \textrm{wTSS} = \textrm{TSS}(\textrm{wCM}), \quad  \textrm{wHSS} = \textrm{HSS}(\textrm{wCM}).
\end{equation}
The weights in the definitions of the wFPs and wFNs allow
mitigating errors such as false positives that precede
the occurrence of an actual positive event and false negatives which are preceded by positive predictions.

\subsection{Operational flare forecasting}\label{sec:operational}
Machine learning theory \citep{DBLP:books/daglib/0097035} points out that training, validation, and test sets should be generated with samples drawn by means of the same probability distribution. However, in the case of flare forecasting this requirement should account for the fact that solar periodicity introduces a bias in chronological splitting. \cite{guastavino2022implementation} introduced an algorithm for the generation of training and validation sets based on proportionality (i.e., training set, validation set, and test set must have the same rate of samples for each sample type described in Subsection 2.1) and parsimony (i.e., each subset of samples must be provided by as few ARs as possible). This algorithm can be exploited in an operational setting if utilized, for example, as follows:
\begin{enumerate}
    \item The current solar cycle is divided into three phases, in which the solar activity increases, reaches its maximum, and decreases, respectively.
    \item Given a time point in the current solar cycle, the corresponding phase is identified.
    \item For the same phase in the previous solar cycle the algorithm computes the rates of the different sample types.
    \item The training and validation sets are generated according to the sample rates from the whole data archive at disposal.
\end{enumerate}
Then the machine/deep learning method is trained by means of the generated training set and the optimal epochs are chosen by means of the generated validation set. When the data set corresponding to the given time point is fed into the trained and validated neural network, flare prediction is performed for the following time point.

\section{Results}
We considered two experiments, both concerning events occurred during Solar Cycle 24 and involving:
\begin{itemize}
    \item The prediction window March-December 2015, which comprises the solar storm occurred in March 2015 \citep{astafyeva2015ionospheric,nayak2016peculiar};
    \item The prediction window January-September 2017, which comprises the solar storm occurred in September 2017 \citep{campi2020machine,qian2019solar}.
\end{itemize}
Since both prediction windows are in the descent phase of Solar Cycle 24, the approach discussed in section \ref{sec:operational} was applied as follows:
\begin{enumerate}
\item A temporal window in the descent phase of the solar cycle and before the prediction window  was identified to compute the sample rates. Specifically:
\begin{itemize}
    \item For the prediction window March-December 2015, the temporal window ranges from 2014-04-30 to 2015-02-28 and the rates of the video samples are $p_\mathrm{X} \approx 0.35\%$,
$p_\mathrm{M} \approx 2.81\%$,
$p_\mathrm{C} \approx 20.53\%$,
$p_{\mathrm{NO1}} \approx 33.25\%$,
$p_{\mathrm{NO2}} \approx 4.56\%$,
$p_{\mathrm{NO3}} \approx 17.89\%$,
$p_{\mathrm{NO4}} \approx 20.61\%$ (where $p_{\mathrm{X}}$ denotes the rate of the $\mathrm{X}$ class samples, $p_{\mathrm{M}}$ the rate of the $\mathrm{M}$ class samples, and so on).
\item For the prediction window January-September 2017, the temporal window ranges from 2014-04-30 to 2016-12-28 and the rates of the video samples are
$p_\mathrm{X} \approx 0.16\%$,
$p_\mathrm{M} \approx 3.18\%$,
$p_\mathrm{C} \approx 16.82\%$,
$p_{\mathrm{NO1}} \approx 38.35\%$,
$p_{\mathrm{NO2}} \approx 5.49\%$,
$p_{\mathrm{NO3}} \approx 15.82\%$,
$p_{\mathrm{NO4}} \approx 20.19\%$. 
\end{itemize}
\item Training and validation sets are generated as in \cite{guastavino2022implementation} by randomly selecting ARs in a temporal interval before the prediction window and by accounting for the rates computed in the previous step. 
\end{enumerate}

In the case of prediction of C1+ flares we labeled the X class, M class and C class samples with $1$ and the other ones with $0$; in the case of prediction of M1+ flares we labeled the X class and M class samples with $1$ and the other ones with $0$. The validation step was realized by selecting the epochs providing the highest TSS and wTSS values, respectively. We found that, in the case of the prediction window March - December 2015 the maximization of the two scores is obtained with the same epoch, which leads to the same confusion matrix and the same skill score values on the test set (see Table \ref{tab:test_March_Dec2015}). On the contrary, in the prediction window January - September 2017 the epochs corresponding to the highest TSS and wTSS values, respectively, were different and the skill score values associated to the choice based on wTSS are significantly higher (see Table \ref{tab:test_jan_sept2017}).

\begin{table*}[ht]
		\centering
		\caption{Results on the test set March 2015 - Dec 2015. The epoch maximizing TSS and wTSS in the validation set is the same. Predictions are performed with this epoch.
}\label{tab:test_March_Dec2015}
{
\begin{tabular}{l | l l l l }
& \multicolumn{2}{c|}{Prediction C1+ flares} & \multicolumn{2}{c|}{Prediction M1+ flares} \\
\cline{2-5}  \\[-8pt]
 & \multicolumn{2}{c|}{Best epoch (TSS/wTSS)} &   \multicolumn{2}{c|}{Best epoch (TSS/wTSS)}  \\
\multirow{2}{*}{
confusion matrix} &  TP=$144$  
& \multicolumn{1}{c|}{FN=$30$} & TP=$30$ & \multicolumn{1}{c|}{FN=$10$} \\
& FP=$136$  
& \multicolumn{1}{c|}{TN=$434$} & FP=$74$ & \multicolumn{1}{c|}{TN=$630$}  \\
\hline\\[-8pt]
TSS &  \multicolumn{2}{c|}{0.589
}  
 & \multicolumn{2}{c|}{0.6449
} \\ 
HSS &   \multicolumn{2}{c|}{0.4861} &     \multicolumn{2}{c|}{0.3676} \\
wFN  & \multicolumn{2}{c|}{32.33}  &     
\multicolumn{2}{c|}{14.25} 
  \\
wFP &    \multicolumn{2}{c|}{158.92} &     \multicolumn{2}{c|}{105.17}  \\
wTSS 
  & \multicolumn{2}{c|}{0.5486
}  
  & \multicolumn{2}{c|}{0.5349
} 
 \\
wHSS &     \multicolumn{2}{c|}{0.4381}  & \multicolumn{2}{c|}{0.2722}   \\
\hline
\end{tabular}
}
\end{table*}

\begin{table*}[ht]
		\centering
		\caption{Results on the test set January 2017 - 30 Sept 2017. In this case the epochs maximizing TSS and wTSS in the validation set are different. Predictions are performed with these two epochs.
}\label{tab:test_jan_sept2017}
\resizebox{0.99\textwidth}{!}
{
\begin{tabular}{l | l l l l l l l l}
& \multicolumn{4}{c|}{Prediction C1+ flares} & \multicolumn{4}{c|}{Prediction M1+ flares} \\
\cline{2-5} \cline{5-9} \\[-8pt]

 & \multicolumn{2}{c}{Best epoch (TSS)} &  \multicolumn{2}{c|}{Best epoch (wTSS)} &   \multicolumn{2}{c}{Best epoch (TSS)} &  \multicolumn{2}{c|}{Best epoch (wTSS)}  \\ 
 
\multirow{2}{*}{
confusion matrix} &  TP=$25$  
& FN=$15$ & TP=$35$ & \multicolumn{1}{c|}{FN =$5$} & TP=$6$ & FN=$3$ & TP=$8$ & \multicolumn{1}{c|}{FN=$1$} \\
& FP=$18$  
& TN=$231$ & FP=$39$ & \multicolumn{1}{c|}{TN =$210$} & FP=$20$ & TN=$260$ & FP=$25$ & \multicolumn{1}{c|}{TN=$255$} \\
\hline\\[-8pt]
TSS &  \multicolumn{2}{c}{0.5527}
 & \multicolumn{2}{c|}{\textbf{0.7184}
}   & \multicolumn{2}{c}{0.5952}
 & \multicolumn{2}{c|}{\textbf{0.7996}
} \\ 
HSS &  \multicolumn{2}{c}{\textbf{0.5358}} & \multicolumn{2}{c|}{0.5295} &    \multicolumn{2}{c}{0.311} & \multicolumn{2}{c|}{\textbf{0.3491}} \\
wFN &    \multicolumn{2}{c}{18.92} & \multicolumn{2}{c|}{6.67}  &    \multicolumn{2}{c}{3} & 
\multicolumn{2}{c|}{1} 
  \\
wFP &   \multicolumn{2}{c}{24.67} & \multicolumn{2}{c|}{56.33} &    \multicolumn{2}{c}{31.92} & \multicolumn{2}{c|}{42.42}  \\
wTSS &  \multicolumn{2}{c}{0.4728}
  & \multicolumn{2}{c|}{\textbf{0.6285}
}  & \multicolumn{2}{c}{0.5573}
  & \multicolumn{2}{c|}{\textbf{0.7463}
} 
 \\
wHSS &    \multicolumn{2}{c}{\textbf{0.4485}} & \multicolumn{2}{c|}{0.4182} &   \multicolumn{2}{c}{0.218} & \multicolumn{2}{c|}{\textbf{0.231}}   \\
\hline
\end{tabular}
}
\end{table*}

We finally focused on the prediction of the two solar storms occurred within the two prediction windows, i.e. the one associated to AR 12297 and the one associated to AR 12673, respectively. 
In both cases, the epochs corresponding to the highest TSS and wTSS lead to the same predictions.

\begin{figure}[h!]
\begin{center}
\includegraphics[width=19.3cm]{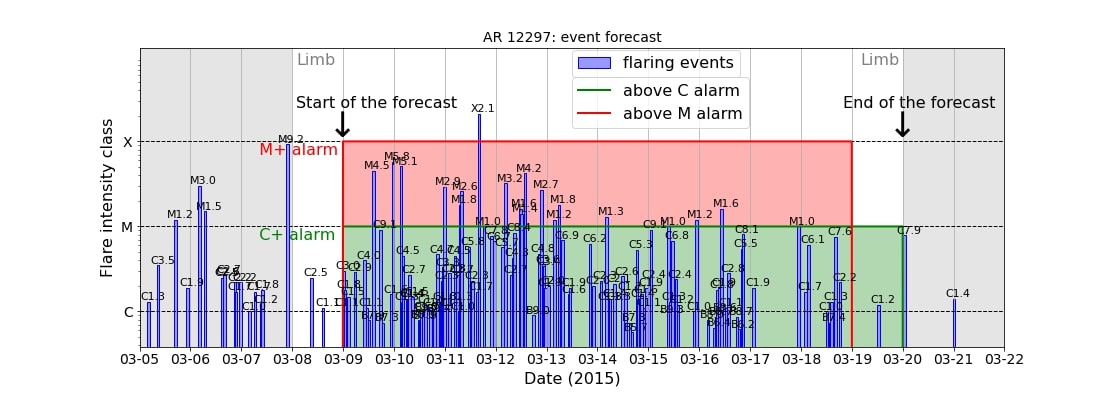}
\end{center}
\caption{Predictions enrolled over time for the solar storm generated by the AR 12297. Warnings of
C1+ flares are in green and warning of M1+ flares are in red lines. The actual flaring events recorded by GOES, together with the corresponding GOES flare
classes are reported. The grey regions correspond to temporal windows out of the field of view of HMI. }\label{fig:March2015_event_forecast}
\end{figure}

Figure \ref{fig:March2015_event_forecast} shows the prediction enrolled over time associated to AR 12297. When the AR started generating C1+ flares it is out from the field of view of HMI, so that the first prediction is made on 2015-03-09 00:00 for the next 24 hours and the last prediction is made on 2015-03-19 for the next 24 hours (after that date the AR was out the field of view of HMI). The samples to predict associated to this AR are 1 X class sample, 8 M class samples, and 2 C class samples. The algorithm correctly sent warnings of the C1+ flares from 2015-03-09 to 2015-03-19 and of the M1+ flares from 2015-03-09 to 2015-03-17. The M1+ flare warning given at time 2015-03-18 is a false positive but, as the figure shows, a M1 class flare had just occurred (the peak time was on March 17th 2015 at 23:34). In Table \ref{tab:test_March2015} the confusion matrices are reported for the C1+ and M1+ flares prediction.

\begin{figure}[h!]
\begin{center}
\includegraphics[width=19.3cm]{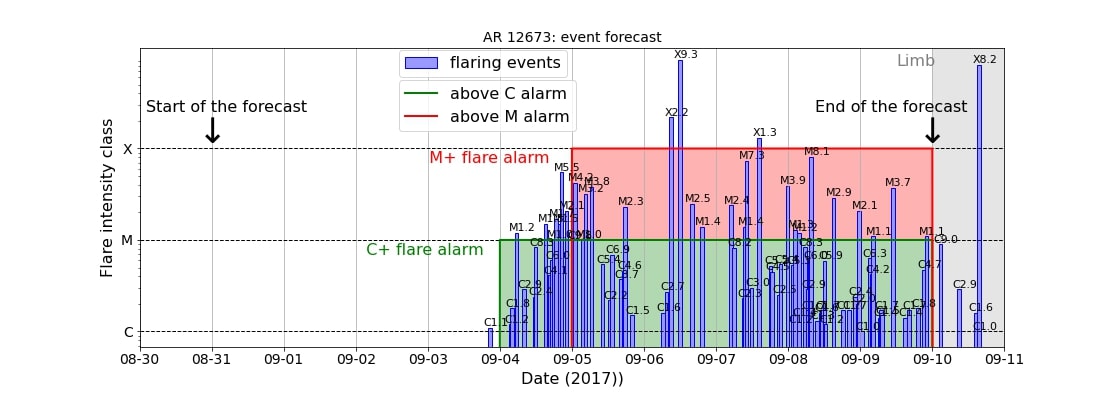}
\end{center}
\caption{Predictions enrolled over time for the solar storm generated by the AR 12673. Warnings of
C1+ flares are in green and warning of M1+ flares are in red lines. The actual flaring events recorded by GOES, together with the corresponding GOES flare
classes are reported. The grey region corresponds to a temporal window out of the field of view of HMI.}\label{fig:Sept2017_event_forecast}
\end{figure}

Figure \ref{fig:Sept2017_event_forecast} shows the prediction enrolled over time associated to AR 12673. The first prediction is made on 2017-08-31 00:00 for the next 24 hours and the last prediction is made on 2017-09-09 for the next 24 hours since after that date the AR was out  the field of view of HMI. The samples to predict are 2 X class samples, 4 M class samples, 1 C class sample, and 3 NO2 class samples. The first 3 NO2 class samples are correctly predicted as negative samples. The algorithm correctly sent warnings of C1+ flares from 2017-09-04 to 2015-03-20 but it missed the C class sample associated to the time range between 2017-09-03 00:00 UT and 2017-09-03 23:59 UT (the figure shows that this C class sample is associated to the occurrence of a C1.1 class flare which has the peak time at 20:50). For the prediction of M1+ flares the algorithm missed the first M class sample associated to the time range between 2017-09-04 00:00 and 2017-09-04 23:59 UT and then from 2017-09-05 00:00 UT it correctly sent warnings until 2017-09-10. In Table \ref{tab:test_Sept2017} the confusion matrices are reported for the C1+ and M1+ flares prediction.


\begin{table*}[ht]
		\centering
		\caption{Predictions the solar storm associated to the AR 12297. The first prediction is made on March 3, 2015 and the last one is on March 19, 2015.
}\label{tab:test_March2015}
{
\begin{tabular}{l | l l l l }
& \multicolumn{2}{c|}{Predictions (C1+ flares)} & \multicolumn{2}{c|}{Predictions (M1+ flares)} \\
\cline{2-5}  \\[-8pt]
\multirow{2}{*}{
confusion matrix} &  TP=$11$  
& \multicolumn{1}{c|}{FN=$0$} & TP=$9$ & \multicolumn{1}{c|}{FN=$0$} \\
& FP=$0$  
& \multicolumn{1}{c|}{TN=$0$} & FP=$1$ & \multicolumn{1}{c|}{TN=$1$}  \\
\hline

\end{tabular}
}
\end{table*}

\begin{table*}[ht]
		\centering
		\caption{Predictions of the solar storm associated to the AR 12673. The first prediction is made on August 30, 2017 and the last one is on September 9, 2017.
}\label{tab:test_Sept2017}
{
\begin{tabular}{l | l l l l }
& \multicolumn{2}{c|}{Predictions (C1+ flares)} & \multicolumn{2}{c|}{Predictions (M1+ flares)} \\
\cline{2-5}  \\[-8pt]
\multirow{2}{*}{
confusion matrix} &  TP=$6$  
& \multicolumn{1}{c|}{FN=$1$} & TP=$6$ & \multicolumn{1}{c|}{FN=$1$} \\
& FP=$0$  
& \multicolumn{1}{c|}{TN=$3$} & FP=$0$ & \multicolumn{1}{c|}{TN=$3$}  \\
\hline
\end{tabular}
}
\end{table*}

\section{Conclusions}

This study shows that the video-based deep learning strategy for flare forecasting introduced in \cite{guastavino2022implementation} can be exploited in an operational setting. This approach populates the training and validation set for the supervised algorithm accounting for the rates of the flare types associated to the specific temporal window of the solar cycle. The prediction algorithm is an LRCN that takes videos of line-of-sight magnetograms as input and provides binary prediction of the flare occurrence as output. The innovative flavor of this approach is strengthened by the use of SOL functions in the optimization step and of value-weighted skill scores in the validation and test phases. The results of two experiments show that this approach can be used as an operational warning machine for flare forecasting on a daily scale.

\section*{Author Contributions}
SG and FM worked at the implementation of the computational strategy. MP, FB, and CC contributed to the formulation of the method and to the design of the experiments. SG, FM, and MP worked at the manuscript's drafting. All authors collaborated to conceiving the general scientific ideas at the basis of the study.

\section*{Funding}
SG acknowledges the financial support of the Programma Operativo Nazionale (PON) "Ricerca e Innovazione" 2014 - 2020. This research was made possible by the financial support from the agreement ASI-INAF n.2018-16-HH.0. 

\section*{Acknowledgments}
The authors enjoyed fruitful discussions with Dr. Manolis Georgoulis, who is kindly acknowledged.


\section*{Data Availability Statement}

The data used for this study named Space-weather
HMI Active Region Patches (SHARP) \cite{bobra2014helioseismic} are provided by the SDO/HMI team, publicly available
at the Joint Science Operations Center. They can be found at the following software repository \cite{monica_g_bobra_2021_5131292}. 

\bibliographystyle{Frontiers-Harvard} 
\bibliography{test}

\end{document}